\pdfoutput=1

\documentclass[graybox]{svmult}

\usepackage{mathptmx}       
\usepackage{helvet}         
\usepackage{courier}        
\usepackage{type1cm}        
\usepackage{amsmath}
\usepackage{makeidx}         
\usepackage{graphicx}        
\usepackage{multicol}        
\usepackage[bottom]{footmisc}

\usepackage{natbib}		

\makeindex             

\def\pbar{\bar{p}}
\def\qbar{\bar{q}}

\def\E{\hbox{E}}

\def\d{\delta}

\def\ovr#1#2{{{#1}\over{#2}}}
\def\dovr#1#2{\ovr{\dd #1}{\dd #2}}
\def\bar#1{\overline{#1}}
\def\dd{{\hbox{\rm d}}}

\begin{document}

\title*{Evolutionary Foundations of Cooperation and Group Cohesion}
\titlerunning{Evolutionary Foundations}
\authorrunning{S.~A.~Frank}
\author{Steven A.~Frank}
\institute{Steven A.~Frank \at Department of Ecology \& Evolutionary Biology, University of California, Irvine CA 92697--2525, USA, \email{safrank@uci.edu}}
%
%
\maketitle

{\hfill\today}\medskip

\abstract{In biology, the evolution of increasingly cooperative groups has shaped the history of life. Genes collaborate in the control of cells; cells efficiently divide tasks to produce cohesive multicellular individuals; individual members of insect colonies cooperate in integrated societies. Biological cooperation provides a foundation on which to understand human behavior. Conceptually, the economics of efficient allocation and the game-like processes of strategy are well understood in biology; we find the same essential processes in many successful theories of human sociality. Historically, the trace of biological evolution informs in two ways. First, the evolutionary transformations in biological cooperation provide insight into how economic and strategic processes play out over time---a source of analogy that, when applied thoughtfully, aids analysis of human sociality. Second, humans arose from biological history---a factual account of the past that tells us much about the material basis of human behavior. [Preprint of article published as: Frank, S. A. 2009. Evolutionary foundations of cooperation and group cohesion. Pages 3--40 in Games, Groups, and the Global Good. S. A. Levin, ed. Springer-Verlag.] } 

\section{Introduction}
\label{sec:1}
People change their behavior in relation to what others do.  The way in which individual behavior changes in relation to others calls upon understanding the evolutionary dynamics of populations.  By ``evolutionary,'' I simply mean the tendency for successful behaviors to increase in frequency.

Understanding the evolutionary dynamics of behavior developed through a long history of study in economics, in game theory, and in evolutionary biology.  The common theme in all fields derives from analysis of self interested actors.  

In economics and game theory, notions of self interest and utility can be problematic; the theory applies to the extent that one accepts certain assumptions about these notions of the individual.  By contrast, a simple measure of self interest arises inevitably in biology from the basic facts of heredity and reproduction: those traits associated with relatively less reproduction have been outcompeted and have disappeared.  Heredity also provides a clear notion of continuity through time, an essential point in the study of behavioral dynamics.  

The clear advantage of biology with regard to the application of evolutionary dynamics led the great statistician and evolutionary biologist R.~A.\ Fisher to say in a 1928 letter to Darwin's son, Leonard:

\begin{quotation}
An engineer finds among mammals and birds really marvelous achievements in his craft, but the vascular system of the higher plants $\ldots$ has apparently made no considerable progress. Is it like a First Law, not a great engineering achievement, but better than anything else \textit{for the price?} Are the plants not perhaps the real adherents of the doctrine of marginal utility, which seems to be too subtle for man to live up to? \citep[p.~94]{Fisher83Bennett}
\end{quotation}

 In other words, evolutionary dynamics of individual interests works beautifully to explain biology, but for humans, the problem appears more complex.  From which, many people conclude that the evolutionary dynamics of self interest teaches us little about humans.  I draw different conclusions.  

On the theoretical side, evolutionary dynamics achieves its greatest development and clarity in biology, because of the clear notions of self interest and continuity through time.  I will therefore develop the evolutionary dynamics of conflict and cooperation within the biological frame, but in a general way that does not depend specifically on biology.  The principles should therefore provide a solid foundation for the application to human behavior.  

On the applied side, understanding the evolutionary dynamics of human behavior is not easy, but should not be abandoned.  Self interest, for all the problems one may wish to raise, remains a powerful theoretical framework in which to analyze human behavior.  Several chapters in this book discuss recent progress and application.  A common view is that, to engineer a social environment that achieves a certain moral goal, such as reduction in hostility or design of fair laws, one must understand the social dynamics in play.  In fact, engineering and dynamics always go together: to control the outcome of a system, one must understand the dynamics of that system.  By this view, evolutionary dynamics and moral engineering are natural partners.

The first part of my paper, on the evolutionary dynamics of conflict and cooperation, provides basic tools that apply across the disciplines of biology, economics, and game theory.  I then turn to a second aspect of evolutionary dynamics: the biological history of evolution.  How has the tension between conflict and cooperation---between individual and group---shaped the history of life?  

One may view this biological history in various ways.  It may be a source of analogy about the dynamical processes that govern human sociality, but similarity arises only through a vague analogy of change in populations.  Or biology may define our history in fundamental ways, because we derive from this history and have been shaped by it in nearly every aspect.  Or, as many humanists prefer, we must maintain a sharp divide between biological history and our understanding of human morality.  

I prefer to think of biological history as both a source of interesting analogy about human affairs and an essential part of our history.  I sketch the connections and the limitations that arise from these lines of thought.  I leave the reader with a series of questions about how much biological analogy and biological history explain modern human sociality.

\section{Scope}
\label{sec:2}

Misunderstanding arises frequently with the words ``evolutionary'' and ``moral.'' I delimit my scope before proceeding.  

I described two distinct meanings of ``evolutionary'' in the introduction.  The first meaning concerns the change in a population over time.  Any economic or game theoretic study that aims to understand human behavior must, at least implicitly, be evolutionary.  At any point in time, each individual has a certain probability distribution over possible behaviors.  As time progresses, each individual's behavioral distribution may change in response to the factors under study.  These simple points alone provide the necessary conditions for the population dynamics of behavior to form an evolutionary system.  From a purely logical or formal perspective, such evolutionary dynamics of behavior do not differ from biological evolution, although the particular rules of continuity and change inevitably differ between particular economic and biological problems.

The second meaning of ``evolutionary'' concerns the specific facts of evolutionary history.  How did humans evolve?  How have our brains been shaped by our past history?  What consequences does that past evolutionary history have for understanding the behavior of modern humans?

With regard to ``moral,'' I consider two distinct positions that parallel the two types of evolutionary analysis.  First, whatever one takes to be the scope of moral studies, most issues concern individual attitudes, beliefs, or behaviors that may change over time in response to those attributes in other members of the population---an evolutionary problem.  Often the issues will turn on some aspect of life that comes down to conflict or cooperation, and what this book labels as the ``global good.''

Second, the particular facts of evolutionary history may help us to understand the dynamics of moral issues.  Such insight may come purely by way of analogy.  For example, the theory of justice introduced in Adam Smith's \textit{Moral Sentiments} and developed by John Rawls forms a very close analogy with one of the fundamental processes that shaped cooperation in biological history \citep{Alexander87Systems,Leigh91individuals,Skyrms96Contract,Frank03cooperation}.  Alternatively, and more controversially, insight may follow from the particular ways in which humans have been designed by natural selection through our evolutionary history.

\section{Evolutionary Dynamics}
\label{sec:3}

I start with the simplest force that favors cooperative evolution, the tendency for similar behaviors to interact.  In the second section I discuss repression of competition, in which reduced opportunities for conflict make cooperation the only way in which to increase payoff.  In the third section I consider correlated interests between individuals, in which an actor places some value on the consequences to those who receive the outcome of the behavior.  In the fourth section I turn to synergism, the positive interaction or feedback between cooperative behaviors with respect to payoff. 

\subsection{Correlated Behaviors and Information about Social Partners}

I start with a simple example to focus the problem.  Suppose a group depends on a common resource---the commons.  That resource may be land that supports farming or a forest that supports wild food products.  The total success of the group depends on the long-term flow of goods from the common resource.  Prudent exploitation maximizes long-term flow and group good; overexploitation reduces long-term flow and group success.  

A self interested individual gains according to two distinct components of success.  First, that individual gains a particular share of the local resource.  Second, the value of the individual's share depends on the total value of the local resource.  

The essential tension of sociality arises from the conflict between an individual's local share and the resource's total value.  An individual always increases its share of the common resource by competing more strongly against neighbors.  However, increased competition leads to over-exploitation of the resource, reducing long-term gain and lowering everyone's success.  Self interested individuals tend to overexploit the common resource, leading to the tragedy of the commons \citep{Hardin68commons}. 

I developed a simple evolutionary model of the tragedy of the commons \citep{Frank94parasites,Frank95groups,Frank98Evolution}.  This model highlights in a clear way the two key forces that can overcome the tragedy of the commons: correlated behaviors between social partners and repression of competition.  In this section, I discuss correlated behaviors in terms of information about social partners.  In the following section, I analyze repression of competition as the second key force that can promote group cohesion and prudent exploitation of shared resources.

\subsubsection{A Simple Model}

Suppose the world is divided into local groups.  Each group has its own common resource, available only to members of that local group.  We seek the behavior adopted by self interested individuals.  We find that behavior by searching for a situation in which, if everyone adopted a particular behavior or nearly so, then no self interested individual could do better by deviating from the population norm \citep{Maynard82Games}.  

I measure individual behavior by the degree to which an individual exploits the common resource.  Values range from 0 to 1. Higher values represent greater individual exploitation; lower values represent more prudent and cooperative individual behavior with regard to the long-term value of the shared resource.  

I seek the population-wide value of behavior, $z^*$, such that anyone who deviates does worse.  To find that value, I measure the behavior of individuals who deviate as $I-z^*=\d$, where $\d$ is a random variable that measures individual deviation.  Each individual lives in a local group.  If we focus on a particular individual within a group, and set that individual's deviation to $\d=x$, then we can use the theory of least squares to write the optimal prediction for group deviation, $G$, given the individual deviation as \citep{Frank98Evolution}
\begin{equation}
	\E(G|\d=x)-z^*=rx.
\end{equation}
We read this as:  given an individual's particular behavioral deviation, $x$, the expected deviation of that individual's group is $rx$, where $r$ is the regression of the average group behavior on individual behavior.  In this case, the regression is equivalent to the correlation between behaviors of members in a group.  This correlation is just a description of pattern without implication about mechanism: we simply note that, given a particular individual deviation, the group deviates to the extent that individual and group behavior are correlated.  Put another way, $r$ measures an individual's information about social partners given the value of the individual's own behavior.  

Individual success depends on the product of two components.  First, measure an individual's share of the local resource as $f(I,G)$, where $I$ is the individual's competitive grab for local share, and $G$ is the local average competitiveness.  The function $f$ rises with $I$ and declines with $G$, because an individual's competitiveness raises its local share, and group competitiveness shrinks its local share.  Second, measure the long-term value of the local resource as $h(G)$, in which long-term value declines as group competitiveness increases.  Thus, we can write individual success as 
\begin{equation}\label{tFitness}
	W=f(I,G)h(G),
\end{equation}
where it would be better to write this expression as the expected payoff given an individual's behavioral deviation, $\E(W|\d=x)$, but for simplicity I just write $W$. 

We analyze how payoff changes with individual behavior by
\begin{eqnarray}
		\dovr{W}{x} &= f_xh + r(fh_y+f_yh) \label{tPartial}\\
		                   &= -C_m + rB_m, \label{hr}
\end{eqnarray}
where subscripts denote partial differentiation with respect to that variable, and $y=rx$ is the group deviation \citep{Taylor96model,Frank98Evolution}.  In Equation \eqref{tPartial}, the first term, $f_xh$, is the marginal change in an individual's share of the local resource as the individual behavioral deviation, $x$, changes.  This measures the direct effect of an individual's behavior on success, holding constant how the individual's behavior correlates with the average competitiveness of neighbors and the value of the group resource.  By convention, we call this direct effect of an individual's behavior on success the marginal cost of cooperation, $C_m$.  In this case, $C_m = -f_xh$, where the minus sign arises because cooperation means a decrease in competitiveness, that is, a decrease in $x$.  

The second term on the right side of Equation \eqref{tPartial} has two parts.  First, $fh_y$ measures the consequences of the marginal increase in the group resource as competition within the group decreases, that is, as the group deviation, $y=rx$, decreases.  Second, $f_yh$ measures the consequences of the marginal decrease in the competitive pressure imposed by neighbors as the group deviation decreases.  By convention, $B_m$ measures the way in which a marginal increase in cooperative behavior among neighbors affects marginal change in individual success.  This marginal benefit term, $B_m$, is weighted by $r$, because group behavior changes at a rate $r$ relative to a change in individual behavior.  Thus, $r$ functions as an exchange rate between the marginal costs of individual cooperative behavior and the marginal gains of group cooperative behavior, rendering the costs and benefits on the common scale of individual payoff.  

The condition for evolutionary dynamics to favor an increase in an individual behavior requires that the change in payoff with an increase in behavioral deviation be greater than zero, that is, $\dd W/\dd x>0$, which also means that
\begin{equation}\label{hrIneq}
	rB_m - C_m > 0,
\end{equation}
an inequality known as Hamilton's rule in biology \citep[see Equation \eqref{hrOrig} below]{Frank06Studies}.  In a moment, I discuss the importance of  $r$, the group correlation.  But first, I look at the outcome of the tragedy of the commons in a very simple case.

In Equation \eqref{tFitness}, let $f=I/G$, which means that an individual's share of the local resource is proportional to its competitiveness, $I$, divided by the average competitiveness of members of the local group, $G$. Let $h=1-G$, which means that the value of the group resource decreases linearly with the average competitiveness of group members, yielding
\begin{equation} \label{tragFitSimple}
	W=\ovr{I}{G}(1-G),
\end{equation}
or
\begin{equation}
	\E(W|\d=x)=\ovr{z^*+x}{z^*+rx}(1-z^*-rx).
\end{equation}
With these assumptions, the behavior that, once nearly adopted by everyone, cannot be improved with regard to individual payoff is
\begin{equation} \label{solnTragedy}
	z^*=1-r,
\end{equation}
obtained by solving $\dd\E(W|\d=x)/\dd x=0$ evaluated at $x=0$, as described in \citet{Frank94parasites,Frank95groups,Frank98Evolution}.   Clearly, as the correlation between social partners, $r$, increases, individual competitiveness, $z$, declines, or, equivalently, individual cooperative behavior increases and enhances the long-term prudent harvesting of the common resource.  

\subsubsection{Interpretation of Group Correlation}

In this model, the severity of the tragedy depends on the behavioral correlation between group members.  If, for example, group members are perfectly correlated, then they all have the same behavioral level of competitiveness, and no one can outcompete a neighbor.  If no gains can be had at a neighbor's expense, then the only way to increase individual gain is by increasing the value of the common good.  As the correlation, $r$, between neighbors declines, opportunity to outcompete neighbors rises, and each individual is favored to raise its competitive efforts even though the outcome is worse for all.  

This model does not assume or depend on any particular mechanism that imposes correlation in the behavior between group members.  In biology, the classical interpretation is that correlated behavior arises from correlated genes, usually between genetic relatives derived from recently shared ancestors.  In a simple case, siblings would be genetically correlated by one-half.  To calculate $r$ in this case, note that, in a group of size $N$, the individual itself composes a fraction $1/N$ of the group, and an individual is correlated to itself by one.  So, for siblings, $r=(1/N)1+[(N-1)/N](1/2)$.  

Correlation does not require genetics and shared genealogy.  Individuals may choose correlated social partners.  Individual choice of where to live may be correlated with behavior, so that those living in a particular place tend to behave in a correlated manner.  Or, there may be some extrinsic force that imposes behavioral correlation.

The notion of information about social partners is very general \citep{Aumann74strategies,Aumann87rationality}.  With genetic relatedness in biology, individuals do not necessarily ``know'' or have direct information about the behavior of their partners.  Rather, if an individual happens to live near correlated individuals, then natural selection will favor those behaviors that exploit the correlations.  By the evolutionary process, the existing correlation becomes exploited as information, and the resulting behavior is shaped in accord with that information \citep{Binmore94Contract,Pollack96strategies,Skyrms96Contract,Frank98Evolution,Frank06Studies}.  In humans, the evolutionary dynamics of behavioral adjustment may be complex.  But, as long as individuals seek self interest by some process of trial and error, they may often come to settle on behaviors that exploit existing correlations: the invisible hand may come to discover and use information about social partners without conscious knowledge of those associations.  Alternatively, direct and conscious information may come into play in some cases.

The point here is that behavioral correlations often shape conflict and cooperation more powerfully than any other process.  The next section turns to mechanisms that may escape the tragedy of the commons when the intrinsic behavioral correlations are low.  In that case, some secondary force must impose correlation to bring individual interests in line.

\subsection{Repression of Competition}
\label{sec:3.2}

The simple tragedy of the commons model in Equation \eqref{tragFitSimple} illustrates well the great importance of behavioral correlation.  Self interested individuals compete at a level $z^*=1-r$, where $r$ is the behavioral correlation among members of a group.  As the correlation declines, competition becomes more severe, and shared resources become over-exploited to the detriment of all.  

If some mechanism creates strong correlations within groups, then self interested individuals naturally adjust their behavior to cooperative ends.  In the absence of intrinsic correlation, behavior tends to be competitive and mutually destructive.  So, in the absence of an intrinsic correlation, what additional force can bring the interests of the competitive group members into line and thereby improve everyone's lot?  

One possibility is that the self interested members of the group would gain by investing some of their own resources in mechanisms that repress competition in their group.  Such policing of competition, by reducing the opportunities for individual gain against neighbors, would have the effect of imposing greater correlation among group members in the payoffs they receive.  As mechanisms that level opportunities for individual gain intensify, individual payoffs become increasingly correlated with other group members independently of the resources that each individual invests in selfish and competitive behaviors \citep[see these references for connections to notions of fairness and justice discussed by Adam Smith and John Rawls]{Alexander79Affairs,Alexander87Systems, Leigh91individuals,Skyrms96Contract,Frank03cooperation}.  

I focus on a simple extension of the tragedy of the commons model from the previous section, in which I add a second behavioral character that determines the extent to which individuals contribute their own resources to repressing selfish, competitive behaviors within their group \citep{Frank95groups,Frank96vary}.

In the previous section, I used a simple payoff function to describe the tragedy of the commons
\begin{equation}
	W=\ovr{I}{G}(1-G),
\end{equation}
where $I$ is the intensity at which an individual competes against neighbors for a share of the local resources, and $G$ is the average intensity of competition within a group.  I extend that model by adding a second behavior expressed by each individual, $A$, the amount an individual invests in mechanisms that police and repress local competition in the group, with $0\le A\le1$.  The average investment in policing per group member is $P$.  With this second character that represses competition, we can now express individual payoff as
\begin{equation}
	W=(1-cA)[P - (1-P)(I/G)][1 - (1-P)G].
\end{equation}
The first term applies a discount to individual success for the cost of investment in the public good through the policing mechanism, where $c$ is the cost per unit investment in policing, $A$.

The second term is the individual's competitive success against neighbors for obtaining a share of local resources:  a fraction $P$ of local resources are distributed evenly to all group members, where $0\le P\le1$ is the average level of investment in the mechanisms that repress competition; a fraction $1-P$ of local resources remains available for splitting by competitive interactions, of which the focal individual acquires its share in proportion to $I/G$, given by the relative competitiveness of an individual, $I$, compared with the average level of competitiveness in the group, $G$.  

The third term quantifies the long-term value of the shared resource.  As before, the resource value declines with local competition.  In this case, $G$ is the average latent competitiveness of individuals, but only a fraction $1-P$ of that competitiveness can be expressed, because local policing represses a fraction $P$ of competitive behavior.  

We need to find the values of the two behaviors, competitiveness and policing, such that when the population adopts the values $(z^*,a^*)$, no individual that deviates can do better.  I discussed the details in \citet{Frank95groups,Frank96vary}; here I give a brief summary.  As before, the correlations in behavior among group members play a key role.  Here, $r_z$ is the correlation in competitive values between a randomly chosen individual and the group average, and $r_a$ is the correlation in the amount invested in policing between a randomly chosen group member and the group average.  

With regard to competitive behavior, self interested individuals are favored to express a level
\begin{equation}  \label{policeZ}
	z^*=\ovr{1-r_z}{1-a^*(1-r_z)}.
\end{equation}
The numerator is the solution for the simple tragedy of the commons model as given in Equation \eqref{solnTragedy}.  The denominator term, $a^*(1-r_z)$, accounts for the amount of competition that is repressed, expressed as uncorrelated behavior $1-r_z$ that is repressed at a level $a^*$; this amount of reduced competition does not lower the long-term value of the shared resource.  In this simple model, competition has a cost only through its affect on the value of the shared resource.  So as mechanisms that repress expression of competition rise, the competitive tendency of individuals also rises.  As I mentioned in \citet{Frank96vary}: 
\begin{quotation}
The high competitiveness in a policing situation is no different from high internal pressure in a fish that lives at great depth. The fish brought to the surface explodes; intense competition and avoidance of repressive policing cause chaos when the same amount of energy is devoted to competition in the absence of repressive policing.
\end{quotation}
We might add an additional direct cost of competitiveness, as in \citet{Frank96vary}, but I do not include that here.  

With regard to repression of competition, self interested individuals are favored to invest in policing the group at a level
\begin{equation}  \label{policeA}
	a^*=\ovr{r_a(1-r_z)-cr_z}{cr_a(1-r_z)}=\ovr{1}{c}-\ovr{r_z}{r_a(1-r_z)},
\end{equation}
with the constraint that $0\le a^*\le1$.  The investment in policing to enforce group cohesion: declines with cost of effective repression, $c$; declines with the intrinsic correlation in competitive behavior, $r_z$, because increased competitive correlation reduces the ability of one individual to outcompete another and thus favors individuals to reduce their competitive tendencies without repression; and rises with the intrinsic correlation in investment in policing, $r_a$, because greater correlation reduces the loss an individual pays relative to neighbors for contribution to policing.  

This simple model captures well how two opposing behaviors together shape the nature of group cohesion.   On the one hand, individual competition within groups inevitably leads to the tragedy of the commons unless check by some opposing force.  Intrinsic correlation in the competitive tendency between group members, $r_z$, can alleviate the tragedy, because correlated group members cannot outcompete their neighbors and so gain by lowering their competitive tendencies.  On the other hand, if the intrinsic correlation in competitive behavior is low, then selfish individuals are often favored to contribute to their own good by preventing the local devastation of their shared resource.  They may accomplish this by investing in mechanisms that repress local competition, such as aspects of policing behavior.  

The two distinct behaviors---individual competitiveness and contribution to group mechanisms that suppress competition---lead to an interesting duality in individual behavior.  The most competitive groups, with low intrinsic correlation, $r_z$, between group members, most strongly favor competitive individuals to contribute resources to the group good through investment in the policing mechanisms.  Increased policing favors individuals to become even more competitive, because competition is often suppressed as fewer shared resources become available for open competition.  So behavioral dynamics tend to favor both greater contribution to policing mechanisms that promote the global good by preserving the shared resource and greater competitiveness of individuals.  Ultimately, the outcome depends on, $c$, how costly it is to develop an effective mechanism to repress competition, and on the intrinsic correlations in behavior that tie the success of individuals to other members of the group.

We may think of repression of competition as a mechanism that enhances local correlation.  In a group that invests $a^*$ to suppress competition, the effective correlation in competitive behavior between group members becomes $a^* + (1-a^*)r_z$.  In words, a fraction $a^*$ of local resources is distributed fairly and without disruptive competition that degrades the common resource, and a fraction $1-a^*$ of local resources remains available for local and destructive competition.  Among that open fraction, $1-a^*$, the intrinsic correlation $r_z$ comes into play, leading to overall correlation in competitive success against neighbors as express by $a^* + (1-a^*)r_z$.

In the models here, I have assumed that each group member begins with the same amount of resources.  Interestingly, if individuals vary in their available resources, even by small amounts, behavior tends to diverge between individuals \citep{Frank96vary}.  The relatively stronger individuals allocate much of their excess resources in policing mechanisms that promote the global good, whereas relatively weak individuals allocate nothing to policing mechanisms that preserve shared resources.   Put another way, small variations in individual resources cause the well endowed to take over social control.

\subsection{Correlated Interests}

The previous sections focused on correlated behaviors.  Such correlation plays a particularly important role when the individuals under study take actions and also receive the consequences of similar actions by others.  Typical games, the tragedy of the commons, and mutual coercion fall into this class in which individuals are both actors and recipients.

In many behavioral situations, an individual acts to affect a recipient, but the recipient does not take any action.  For example, an individual may provide aid to a brother or offspring without reciprocation.  Theories based on correlated behaviors do not apply in these sorts of one-sided interactions.  So, how may we account for altruistic behaviors in these cases?

Presumably, an actor who makes a costly behavior in favor of a recipient must value the recipient's interests.  In biology, we have an extensive theory by which we can calculate how much an actor values different recipients \citep{Frank98Evolution}.  In this case, individuals do not consciously put different values on different recipients.  Instead, natural selection shapes the behaviors of actors in relation to different recipients.  Outside of a biological framework, no theory provides an absolute basis for assigning relative values.  Because I focus in this section on general forms of the theory that transcend biology, I limit my discussion here to how we may describe relative valuation between actors and recipients, without regard to what causes such valuations.

Suppose that, in valuing the total payoff to an individual in return for some behavior, $z$, we consider the individual's relative regard for others and self as
\begin{equation}
	W(z) = vW_o(z)+W_s(z),
\end{equation}
where $W_o$ is the valuation of others affected by an actor's behavior,  $W_s$ is the valuation to self as a consequence of an actor's behavior, and $v$ is an exchange rate between self valuation and valuation of others in a particular behavioral situation.   The change in an actor's total valuation in return for a small change in behavior can be written as $\dd W/ \dd z$, and, using primes to denote differentiation with respect to $z$, we may write the condition for an increase in the particular behavior to be favored as
\begin{equation}
	W'=vW'_o+W'_s>0.
\end{equation}
It is often useful to write this condition equivalently as
\begin{equation}  \label{hrOrig}
	vB_m - C_m > 0,
\end{equation}
where $B_m$ is the marginal benefit to the recipient, and $C_m$ is the marginal cost to the actor.  The behavior is favored when the value-weighted marginal benefits to others are greater than the marginal costs to self.  We could apply this method of valuation to any sort of game or economic analysis of self interest.  

In biology, the condition in Equation \eqref{hrOrig} for a behavior to be favored by natural selection is known as Hamilton's rule \citep{Hamilton70model}.  The identical form of Equation \eqref{hrIneq} misleads \citep{Frank98Evolution,Frank06Studies}.  In Equation \eqref{hrIneq}, $r$ measures the effect of behavioral correlation between neighbors on the direct success of the actor.  The correlation may arise by genetic similarity, but other processes that impose correlation work in the same way.  By contrast, $v$ in Equation \eqref{hrOrig} measures an actor's regard for the success of a recipient of the behavior---often, the recipient does not express any behavior in return and has no direct affect on the actor.  Biology values $v$ by the genetic similarity of the actor to the recipient.

\subsection{Synergism and the Origin of Mutually Beneficial Behaviors}

Different groups with complementary skills or resources can achieve synergistic gains by cooperating.  However, if few tend to join cooperative ventures, then an individual who puts forward its potentially complementary resource may end up losing that resource.  By contrast, if everyone tends to join synergistic activities, then no one gains by withholding their complementary skill, and cooperation is easily maintained.  In this case, the difficulty concerns how to start synergistic partnerships which, once they become common, are easily maintained by advantages to self interested individuals \citep{Axelrod81cooperation}.  

Much of the cooperative structure of life in biology and in human behavior arises from such synergistic interactions.  The positive feedbacks and consistency of cooperation often become so deeply embedded that their very existence can be difficult to discern.  The more cooperative and nonvarying the interaction, the less one tends to notice it.

For example, many animals depend on the numerous bacteria that they carry in their bodies: the bacteria provide essential dietary products to the host.  The bacteria, in turn, sometimes cannot live without their hosts.  This modern synergism is easy to understand: mutual dependence and mutual gain, although the potential for conflict remains within the alliance.   In humans, specialized production and trade engenders mutual dependence and enhanced alliance; subsequent conflict runs within the constraints of synergistic benefits.  

How do transitions occur between an initially uncooperative situation and a final situation in which cooperation reigns?  Often, in the initially uncooperative state, no one gains by offering their special skills or resources if the behavioral or structural situation does not return synergistically matching skills or resources.  So, the difficulty is how to get things started.  

The transition from an initially uncooperative state to a cooperative one often turns on the behavioral correlation between potential partners \citep{Axelrod81cooperation}.  For example, if cooperation is rare, but the behavioral correlation is high, then those rare individuals who tend to cooperate will often meet cooperative partners, and so mutually beneficial synergism can get started.  As before, the cause of such correlation does not matter: cooperative individuals may be able to recognize each other and seek each other, or the few cooperative individuals may for accidental reasons tend to live near each other.

A simple game captures the way in which behavioral correlations influence transitions from uncooperative beginnings to synergistically cooperative and mutually beneficial social structures \citep{Frank98Evolution}.  Consider, for example, a particular interaction between an individual from group I and an individual from group II.  

\begin{figure}
\sidecaption
\includegraphics[scale=1]{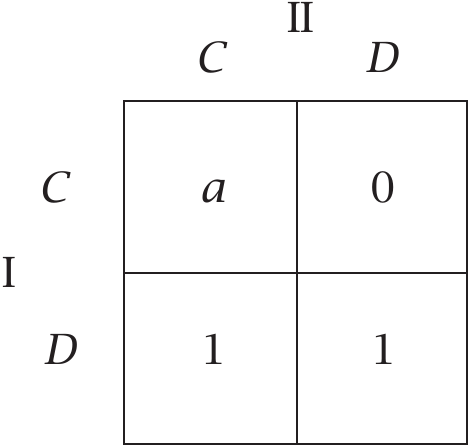}
\caption{Matrix for a two-player game. The cells show the payoff to player I given strategies by two players in an encounter. The $C$ and $D$ strategies correspond to cooperation and defection. The payoff to player II in this symmetric game can be obtained by transposing the matrix.  I assume $a>1$.}
\label{fig:saddle} 
\end{figure}

The game matrix in Figure \ref{fig:saddle} shows the payoffs for three different outcomes.  First, an individual keeps the initial resource if it does not enter a joint venture with a potential partner.  In the figure, withholding cooperative behavior is the D or defect strategy.  A defector keeps the initial resource, in this case equal to a payoff of one unit, no matter what the partner does.  Second, if an individual puts forward its resource in the C or cooperative strategy, and the partner does not reciprocate, then the individual loses its resource and receives a payoff of zero.  Third, if both cooperate, then both gain the synergistic benefits with a payoff of $a>1$.  

In a particular encounter, player I cooperates with probability $p$, and player II cooperates with probability $q$ (mixed strategies allowed).  The payoff to player I is
\begin{equation} \label{saddleGame}
	w(p,q) = 1 + p(aq-1),
\end{equation}
and the payoff to player II is $w(q,p)$ by the symmetry of the payoff structure.  Here, the players are drawn from separate populations, with average strategies $\pbar$ and $\qbar$, respectively.  

From Equation \eqref{saddleGame}, we can see that player I is favored to increase its level of cooperation, $p$, when, on average, 
\begin{equation}  \label{saddleCond1}
	q>1/a.
\end{equation}
To understand this condition, we must consider what information player I has about the expected behavior, $q$, of its partner, player II.

Player I has information about player II's strategy to the extent that interacting pairs have correlated behaviors.  Suppose $p$ deviates from its population average by $\d=p-\pbar$.  Then we can describe the information player I has about the expected behavioral deviation of its partner by a regression equation
\begin{equation}
	\E(q|p)-\qbar = r\d,
\end{equation}
where $r$ is a regression coefficient, because the players are drawn from different populations \citep{Frank94species}.  Given this regression equation, we can express the expected value of player II's behavior given information about player I's behavior as $\qbar+r\d$, and so the condition in Equation \eqref{saddleCond1} for player I to be favored to increase its cooperative behavior becomes
\begin{equation}
	\qbar + r\d > 1/a.
\end{equation}
If we start with the absence of cooperation, $\pbar=\qbar=0$, then full cooperation can spread only when 
\begin{equation}
	r>1/a. 
\end{equation}
Thus, a significant behavioral correlation is needed to make the transition to a cooperative state.  Once the populations have moved to full cooperation, they gain from synergistic benefits because $a>1$.  At that point, no correlation is needed to maintain cooperation.  Thus, correlation drives the initial transition, but does not play a role in subsequent maintenance.  

With full cooperation, each population may become dependent on the skills and resources of its partner population.  At that point, mutual dependence causes cooperation to become essentially irreversible \citep{Frank95symbiosis}.  Much of cooperation and the evolution of social structure may follow such a path, through which brief periods of information about social partners allow mutually beneficial traits to flourish.  As those traits flourish, they become embedded in the structure of opportunities available and payoffs gained.  Such traits of mutual dependence may come to seem more as fixed aspects of the social environment than as interesting characteristics reflecting the social tension between cooperation and conflict.

\section{Biological History}

Some people may believe that biological history can teach us little about our own species' conflicts, cooperative associations, and moral dilemmas.  But, upon study, one has to be surprised by how often the basic forces of social tension in human life have deeply and inexorably shaped biological history.  The lessons drawn from such similarity are, certainly, points of debate.  But before we can consider the debate, we need some facts to set a common ground for discussion.

I organize biological aspects of cooperation along the lines of the four major forces that shape the nature of conflict and cooperation among self interested individuals.  Those forces are correlation between social partners, repression of competition, correlated interests between actor and recipient, and synergism.  In this section, I consider the role played by each of those forces in the history of life.  The broad topic is, of course, too great to cover fully.  So I use a few examples to illustrate the key points.  

\subsection{Correlated Behavior and the Tragedy of the Commons}

The tragedy of the commons arises because self interested individuals gain by competing against neighbors.  Rapacious individuals outcompete their neighbors and gain a larger share of the local resource.  But rapacious behavior depletes the local resource in a way that reduces the long-term yield, causing harm to all members of the group.  A mechanism that causes correlation in behavior between group members favors prudent behavior, greater sustainable productivity, and benefit to all group members.  

I applied the tragedy of the commons to problems in biology \citep{Frank95groups,Frank96virulence}, extending a long history of work on group selection \citep{Hamilton67ratios,Hamilton72insects,Hamilton75Anthropology,Lewontin70selection,Leigh77group,Leigh91individuals,Wilson80Communities,Colwell81ratios,Szathmary87life}.  My work arose from study of biological problems such as the sex ratio of progeny produced by mothers within small isolated groups and the amount of harm a parasite causes to its host.  I discuss how these applications and related problems have grown into a major field of study in evolutionary biology \citep{Rankin07biology}.

\subsubsection{Sex Ratio: the Production of Males as a Competitive Trait}

In almost all animals, females produce babies and males do little except compete, mate, and provide sperm.  With regard to reproduction, females are productive and males are competitive.   In some animals, males contribute more that just matings, and the situation is more complex.  But the vast majority animals follow the simple dichotomy.  For example, one can think of most male insects as the mother's competitive winged sperm.

Consider the male-female distinction from a mother's point of view.  She can make a daughter, who produces babies.  Or she can make a son, who competes with other males for matings but produces nothing directly. 

A mother's investment in sons is an investment in a trait to compete against other mothers in the local mating group \citep{Hamilton67ratios}.  The more a mother invests in sons, the more grandchildren she will have through her sons.  Those extra grandchildren come at the expense of reduced numbers of grandchildren through sons by other mothers in the group, because the total number of grandchildren is fixed by the number of productive daughters that are made.  

This problem of allocation of resources to competitive sons is formally equivalent to the tragedy of the commons problem that I introduced earlier in Equation \eqref{tragFitSimple} \citep{Frank06Studies}.  In particular, suppose a mother allocates a fraction $I$ of her reproductive resources to sons and a fraction $1-I$ to daughters.  Assume that, in a local group, daughters and sons grow up and mate with each other, and then the mated daughters disperse to find new reproductive opportunities---a pattern of mating and dispersal that occurs in many insects \citep{Hamilton67ratios}.  In a local group, suppose that, on average, mothers allocate $G$ of their resources to sons and $1-G$ to daughters.

A mother's payoff is the combination of her success through her sons, $W_s$, and her success through her daughters, $W_d$, yielding total success as $W=W_s+W_d$.  For sons, the payoff follows exactly the tragedy of the commons expression for payoff in Equation \eqref{tragFitSimple} as
\begin{equation}
	W_s=\ovr{I}{G}(1-G).
\end{equation}
The first term, $I/G$, accounts for the relative success of a mother through the matings obtained by her sons.  For example, a mother may make $K$ babies of which a fraction $I$ are sons, giving her $KI$ sons.  In the group, there are $N$ mothers who, on average, each make $KG$ sons.  So the fraction of all males in the group by the focal mother is $I/(NG)$.  Those males compete for matings among the local resource, the $KN(1-G)$ daughters produced by all mothers in the local group.  Combining the terms and dropping $K$ as an arbitrary proportionality constant gives $W_s$.  Success through daughters is the number of daughters produced by a mother, $K(1-I)$, and again we drop $K$ as a proportionality constant.  Combining success through sons and daughters yields total payoff as \citep{Hamilton67ratios}
\begin{equation}
	W=\ovr{I}{G}(1-G)+1-I.
\end{equation}

We apply the same methods used to obtain Equation \eqref{solnTragedy}.  The solution is $z^*=(1-r)/2$, where $z^*$ is the fraction of resources a mother allocates to sons such that, if nearly everyone adopts this behavior, no behavior that deviates from it can obtain a higher payoff, and $r$ is the correlation in the sex ratio produced by mothers within a local group \citep{Frank85wasps,Frank98Evolution}.  

The solution parses more easily when we write the best allocation to sons and daughters as a ratio $1-r:1+r$.  Here, the term $1-r$ for male allocation arises from the equivalence between male allocation and the tragedy of the commons \citep{Frank06Studies}.  Sons are the direct expression of a mother's competition against neighboring mothers. Sons are made at the expense of daughters. Daughters contribute to the success of all mothers in the group by providing mates for those mothers' sons.  

The term $1+r$ for the relative value of female allocation arises as follows.  Each extra daughter made by a mother contributes to the mother's success directly through the grandchildren produced by the daughter---the valuation of one.  In addition, an extra daughter provides an extra mate to males in the local group.  That extra mate for sons accrues to the strategy pursued by the focal mother in proportion to the correlation between the mother's strategy (her sex ratio) and the average strategy in the local group \citep{Frank98Evolution}.

This sex ratio model makes a simple qualitative prediction.  As the correlation between mothers declines, mothers compete more intensely with each other by raising their relative allocation to sons.  If we assume that behavioral correlation in a small group arises mainly from a mother's correlation to herself, then in the simplest case $r=1/N$, where $N$ is the number of mothers in the local group.  This gives us Hamilton's \citeyearpar{Hamilton67ratios} famous model of the sex ratio under local competition for mates.  With this expression, $r$ declines as the number of mothers in the group, $N$, rises.  So the prediction becomes: as $N$ rises, the fraction of males produces by mothers should rise.

Many studies show that as more mothers contribute to a local group, the competitive allocation to sons rises \citep{Godfray96studies,Hardy02Methods}.  My own study on fig wasps provides a simple and direct demonstration \citep[see also Herre \citeyear{Herre85wasps}]{Frank85wasps}.  In the species I studied, a mated fig wasp female gets inside a fig, lays her eggs, and dies.  More than one mated female may lay her eggs within a fig during a short window of a few days.   About four weeks later, the male offspring emerge first within the dark cavity in the center of the fig.  The males  mate with the quiescent females.  After a few days, one of the males chews a tunnel through the wall of the fig, stimulating the mated females to emerge, exit, and find another fig to start the cycle anew.  

\begin{figure}
\sidecaption
\centerline{\includegraphics[width=4.5in]{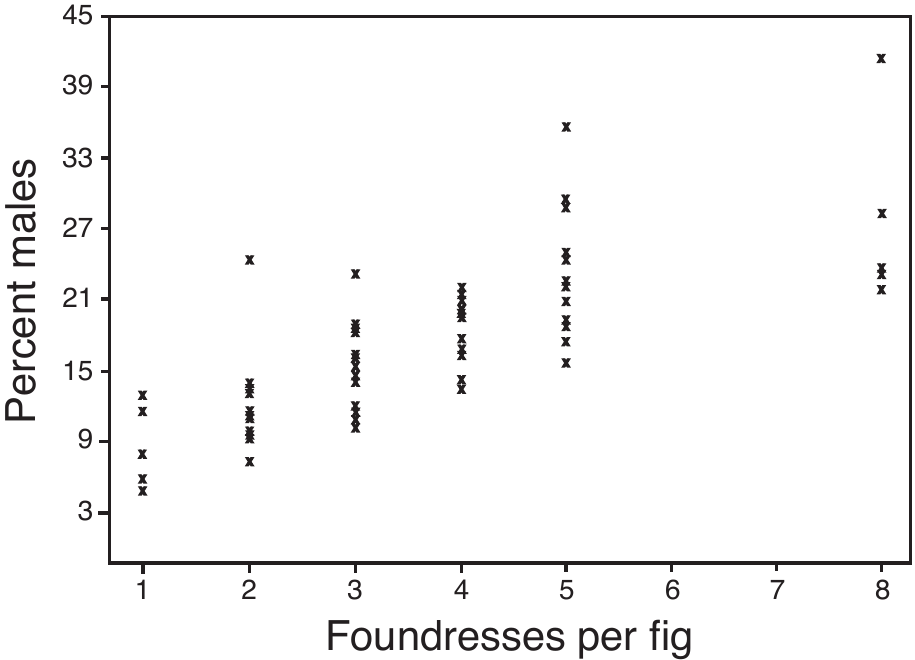}}
\caption{Sex ratio of fig wasps.  Foundresses per fig represents the number of mothers laying eggs in each fig.  The numbers of foundresses were controlled experimentally.  Each ``X'' marks the sex ratio of all foundresses in a single fig.  The percent males rises with the number of foundresses, matching the tragedy of the commons prediction that mothers increase their competitive allocation to sons in response to an increase in the number of other mothers that compete in the local group. Redrawn from \citet{Frank85wasps}. }
\label{fig:figSR} 
\end{figure}

Fig biology imposes exactly the life course assumed by the sex ratio model of local mating and competition among males \citep{Hamilton79Insects}.  To test the theory, I manipulated the number of mothers that enter each fig.  Do these tiny mothers, each less than 2mm, detect the number of other mothers in the dark fig cavity and adjust their allocation to competitive sons?  Figure \ref{fig:figSR} shows that they do.  

\subsubsection{Parasite Virulence: A General Model for Prudent versus Rapacious Exploitation of Resources}

Some parasites exploit their hosts in a prudent way, taking the resources that they need without causing noticeable damage. Prudent exploitation yields sustainable benefits to the parasite as long as the host remains healthy. Other parasites attack their host more quickly and vigorously. Rapid exploitation may allow the parasites to achieve higher reproductive rates, but damage to the host reduces the parasites' opportunity for sustainable yield \citep{Frank96virulence}.

Following this economic line of thought, each parasite faces a tradeoff when increasing the rate at which host resources are used.  Greater exploitation has the benefit of more rapid reproduction and transmission to new hosts, but carries the cost of reducing the host's ability to procure more resources in the future. For each host-parasite interaction, there may be a particular optimum schedule of host utilization that maximizes the parasites' balance between rapid transmission and the time before the host dies \citep{Fenner56mosquitoes,Levin81systems,Anderson82parasites,Levin83Coevolution}.

One process missing from the tradeoff between transmission and virulence concerns the ``social'' aspect of parasite interactions. Suppose that prudent exploitation of a host maximizes a parasite's reproduction. Natural selection then favors each parasite, when alone in a host, to follow the prudent strategy. There is, however, a problem when two or more parasites with different strategies occupy the same host. If one strategy extracts host resources rapidly and reproduces quickly, then the host may die in a short time. A prudent strategy would have relatively low reproduction when paired in a host with a rapacious strategy because, for both strategies, the host is short-lived, and the rapacious strategy reproduces more rapidly than the prudent one.  This is the tragedy of the commons.

\begin{figure}
\sidecaption
\centerline{\includegraphics[width=3.0in]{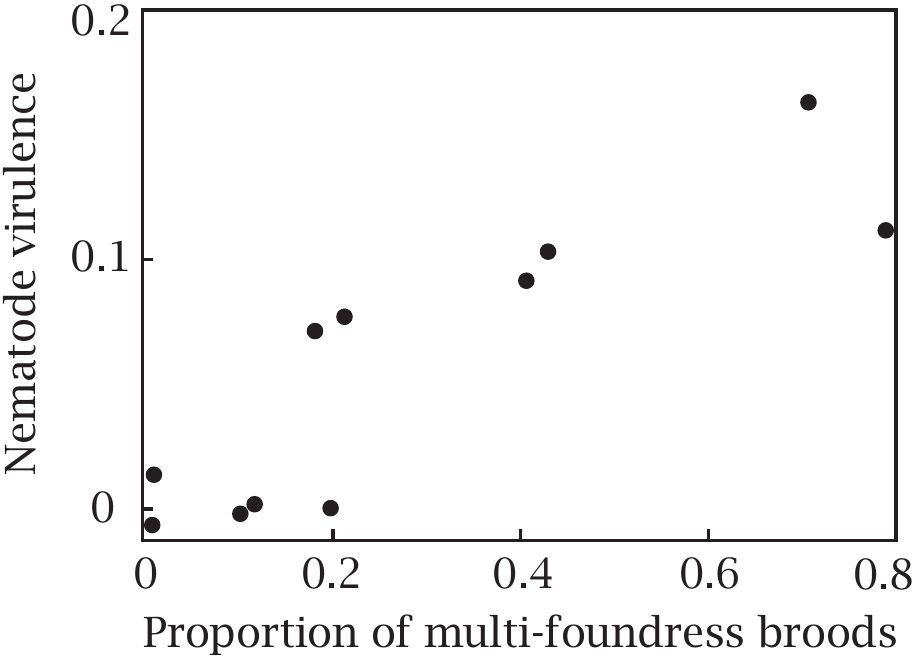}}
\caption{Virulence of nematodes infecting fig wasps.  The fig wasps I described in the sex ratio section often carry parasitic nematodes.  \citet{Herre93wasps} studied how much harm these parasitic nematodes cause to their hosts.  He predicted that greater mixing between nematode lineages would reduce the correlation of behavior (relatedness) within hosts and lead to greater virulence.  The data support the prediction.  Here, multi-foundress broods measure the fraction of figs in which more than one wasp entered.  The more often multiple foundresses enter a fig, the more often the nematode lineages will likely mix, reducing within-host correlation.  Herre measured virulence by $1-f_i/f_u$, where $f_i$ and $f_u$ are the number of babies produced by infected and uninfected wasps, respectively.  Lower productivity of infected wasps corresponds to higher virulence. Redrawn from \citet{Herre93wasps}.}
\label{fig:herreVir} 
\end{figure}

Correlation in behavior between members of a group mitigates the tragedy of the commons.  In biology, the correlation typically arises by genetic relatedness within a group.  In the case of parasite virulence, the prediction is that more related parasites within a host will behave more prudently, competing less intensely among themselves and causing less harm to the host \citep{Hamilton72insects,Bremermann83virulence,Frank92virulence,Frank96virulence}.  Figure \ref{fig:herreVir} supports the predicted trend: reduced correlation among parasites increases the damage caused to the host.  

The pattern in Figure \ref{fig:herreVir} leaves open the issue of whether competition between unrelated parasite lineages plays a direct role in causing harm to the host. \citet{de05infections} showed that, in the parasite that causes malaria, the more competitive parasite lineages did outcompete other parasites within the host and did cause greater harm to the host.  This study of the malaria parasite ties the direct competition over local resources to the harm caused to the public good---the health of the host that provides resources to the parasites.

The problem of parasite virulence captures well the essence of many biological examples of the tragedy of the commons \citep{Frank96virulence}.  For example, the most primitive cells probably contained several molecules that could make copies of themselves---the primitive genes.  Each trait of those replicating molecules was selected according to the balance between individual benefit from rapid exploitation of local resources and group benefit from prudent exploitation of local resources. In other words, the problem of cooperation versus conflict in groups arose in the earliest stages of biological history.  

\subsubsection{Scale of Competition and the Role of Group Productivity}

I presented a simple tragedy of the commons model in Equation \eqref{tragFitSimple}.  In that model, individuals can potentially gain by restraining competition in order to enhance the productivity of their group.  Individuals in more productive groups benefit by getting a piece of a greater local resource, even if their piece may be smaller than their neighbors'.   

This simple model for the tragedy of the commons makes implicit assumptions about how individuals compete and how we measure success.  In evolutionary models, we generally measure the success of an individual relative to some base population, because we are interested in whether a behavior is gaining or losing in frequency in response to its success relative to other behaviors in the comparison population.  

Suppose we regard the local group as part of a broader population, and we measure the success of each individual relative to the broader population.  Then a cooperative individual can gain in the population by trading a smaller share of the local resource in return for a greater total value of the local resource.  In other words, a prudent group gains greater total productivity, benefiting all members of the group.  The interpretation of differential group productivity benefiting members of prudent groups matches the tragedy of the commons model that I gave earlier.  In this case, an individual's success is measured relative to the broad population composed of many local groups.

But what if an individual's success is measured relative only to other members of its local group?  For example, if there is no competition between members of different groups, then each individual's relative success arises only from its advantage or disadvantage compared to its neighbors in its local group \citep{Wilson92populations,Taylor92approach,Taylor92environment,Queller94populations}.  

We can, in general, define the problem by the scale of competition \citep{Frank98Evolution}.  Let the scale of competition, $s$, be the probability that an individual ultimately competes against and measures its relative success against only local group members, and $1-s$ be the probability that an individual ultimately competes against and measures its success against members of the broader population.  The base population for measuring success determines how a particular level of success translates into change in the frequency of a behavioral strategy.

With this definition for the scale of competition, we can extend the tragedy of the commons model in Equation \eqref{tragFitSimple} to
\begin{equation}
	W=\ovr{I}{G}\Bigg(\ovr{1-G}{s(1-G)+(1-s)(1-z^*)}\Bigg),
\end{equation}
where $z^*$ is the average level of competitive behavior in the population, $I=z^*+\d$ is the deviation from the average by a randomly chosen focal individual, and $G=z^*+r\d$ is the deviation of the group average by the focal individual's local group.  By following the approach given earlier, we can find the value of $z^*$ that, once adopted by the population, cannot be beat.  Because $z^*$, the level of competitiveness of individuals, varies in this model between zero and one, we can write $1-z^*$ for the level of individual cooperation, yielding
\begin{equation} \label{scale}
	1-z^*=r\Big(\ovr{1-s}{1-rs}\Big),
\end{equation}
where $r$ is the correlation in behavior within local groups.  Increased behavioral correlation, $r$, favors cooperation.  By contrast, increased local competition, $s$, reduces cooperation.  An individual cannot gain by providing benefit to a neighbor if the individual's ultimate success is measured only against neighbors.  Cooperation can increase only to the extent that an individual ultimately competes against and measures success against members of other groups.

\citet{Griffin04bacteria} studied how behavioral correlation and the scale of competition jointly determine cooperative behavior in bacteria.  Pathogenic bacteria often face iron limitation when living within a host; hosts often withhold iron as a defense against bacteria.  Some bacteria can secrete a molecule---a siderophore---that scavenges iron from the host.  The bacteria then take up siderophore-iron complexes to overcome their deficiency.  

Siderophore production is a public good: costly for individuals to produce and equally beneficial for all members of the local group.  In particular, any member of the local group can take up a siderophore-iron complex independently of who originally secreted the siderophore.

Griffin et al.\ experimentally varied behavioral correlation by changing the amount of mixing between different bacterial clones.  Relatively pure clones cause high behavioral correlation through genetic similarity. Mixed clones have lower behavioral correlation because of greater genetic diversity.  They varied the scale of competition by altering the constraint placed on the contribution of local groups to the following generation.  If all groups contribute equally by constraint, then individuals compete only locally within their group, and the scale of competition is entirely local.  If groups contribute in proportion to their productivity, then individuals compete fully with members of other groups, and the scale of competition is global.  

\begin{figure}
\sidecaption
\centerline{\includegraphics[width=4.0in]{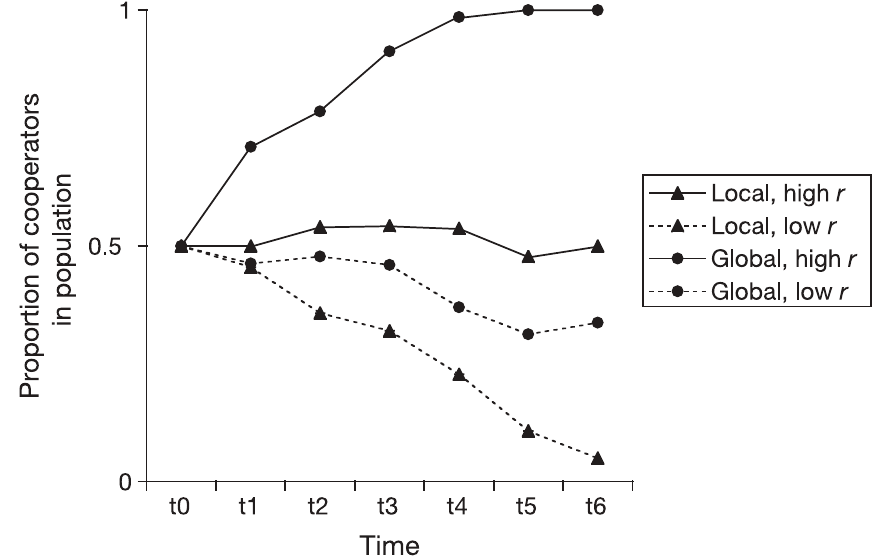}}
\caption{Evolutionary change in siderophore production.  Time moves from left to right, each unit representing one day and one round of mixing of the bacteria to impose either local or global competition.  The behavioral correlation, $r$, was controlled by the degree of mixing between different bacterial clones.  From Figure 3 of \citet{Griffin04bacteria}.}
\label{fig:siderophore} 
\end{figure}

The experiment set the conditions that determine behavioral correlation and the scale of competition.  Then, over time, the evolutionary change in populations was followed with regard to siderophore production, which measures production of a public good and the degree of local cooperation.  Figure \ref{fig:siderophore} shows that the experiment supports the predictions of Equation \eqref{scale}: greater behavioral correlation and global competition increase cooperation.

\subsection{Repression of Competition}

Correlated behaviors align interests and favor reduced competition.  But highly cooperative behavior often occurs in nature with little correlation in the intrinsic tendency of individuals.  To achieve that high level of cooperation in the absence of intrinsic correlation, there must be some force extrinsic to each individual that tends to align interests and behaviors \citep{Frank03cooperation}.

Reduced opportunity for competition can align interests.  If an individual cannot compete against neighbors, then that individual can increase its own success only by enhancing the efficiency and productivity of its group \citep{Leigh77group,Alexander79Affairs}.

In the first example, a randomization process assigns success to individuals independently of their behavior within the local group, preventing any individual from gaining by competing against neighbors.  In the second example, powerful individuals within the local group repress competition between lower ranking individuals; such policing of competition appears to play a key role in social integration within the group.  In the final example, one species domesticates and essentially enslaves another species.  The master species gains by preventing competition between enslaved members of the group, in order to prevent wasted energy devoted to internal competition.

\subsubsection{Randomization and Fairness}

Sexual reproduction mixes the genes from two parents to make an offspring.  Each parent contributes one half of the genes.  Meiosis is the process by which each parent selects one half of its own genes for transmission to the child.  Biologists often refer to the process as ``fair meiosis,'' to emphasize that each copy of a gene has an equal chance of being chosen.  This randomization process means that no gene copy can gain an advantage over other gene copies in being transmitted to the offspring.  

With no opportunity for local competition, all gene copies gain only with the enhanced success of the whole group \citep{Leigh71Diversity}.  In this case, we call the group of the genes the ``genome.''  The unity of the genome, and thus the unity of the individual, is so nearly complete that one often thinks of the genome in a unitary way rather than as a collection of cooperating genes \citep{Maynard95Evolution}.  However, competition between gene copies does occur in nature, in which one gene copy increases its chance of getting transmitted to the offspring at the expense of other gene copies \citep{Crow79rules}.

Competition between gene copies reminds one that the reproductive fairness and the near unity of the genome evolved in the face of competitive pressure between neighbors.  The puzzle concerns how the process of natural selection, acting on the interests of the individual gene copies, led to particular biochemical mechanisms that typically repress internal competition \citep{Frank95groups}, and how those mechanisms can sometimes be subverted by certain competitive types.

This example shows that the very foundations of sex, reproduction, and the genetic transmission of information arose from the group cohesion of a collection of genes.  That cohesion was created by processes that repress internal competition and bind the interests of the separate genes to the group \citep{Maynard95Evolution}.  

As I mentioned, the cohesion of the genome often appears so complete that we use the word ``individual'' to refer to a single genome---a single collection of genes in an organism.  But that ``individual'' is a constructed group, cohesive only because of the high reliability of the internal mechanisms that ensure reproductive fairness among gene copies.  

The occasional gene copies that subvert these fairness mechanisms and outcompete neighbors emphasize that unity in biology must always be constructed and maintained.  Such unity arises solely from self interested actors.  The system of cohesion built by those self interested actors must enforce against the competitive tendencies of those same actors.  And so it always goes: conflict and cooperation in constant tension and never separable.

\subsubsection{Policing and Repression of Competition}

Individuals in a group may prevent others from competing.  Such repression of competition by third-party policing reduces opportunity for individuals to gain at the expense of their neighbors.  Once again, in the absence of opportunity to outcompete a neighbor, an individual can increase its own success only by enhancing group efficiency and productivity.  

Policing of competition can be a very effective mechanism to promote group cohesion.  However, policing competition between others can be dangerous or costly.  Why should a self interested actor take on the costs of the policing role?  In Section \ref{sec:3.2}, I showed the conditions under which an individual may gain more by its benefit from living in a more productive group than it loses by the costs of policing.   I also mentioned how the theory predicts an interesting asymmetry with regard to policing:  those individuals with relatively greater vigor or resources are favored to take on the policing role, whereas those with relatively lower vigor or resources do not gain from policing \citep{Frank96vary}.  Thus, the theory predicts that the relatively powerful individuals impose social control on the group when effective mechanisms exist for dominant individuals to repress competition and promote group cohesion.

\citet{Flack05study,Flack05species} studied policing and group cohesion in pigtailed macaques.  Dominant males intervene to control disputes between pairs of lower ranking individuals.  Those policing acts usually do not favor one competitor over the other, but rather the intervention puts an end to the conflict.  

\citet{Flack05study,Flack05species} analyzed the consequences of policing interventions for various aspects of group cohesion.  They compared two situations in a semi-natural captive colony.  In the baseline case, the dominant males were present and acted in their normal way to settle disputes.  In the ``knockout'' experiment, Flack et al.\ removed the dominant males and placed them just outside a wall that bounded the colony.  The dominant males were visible to the colony members but could not intervene.

When policing males were removed, the amount of aggressive behavior in the colony increased.  Measures of aggression included initiation of conflicts, intensity of conflicts, biting, and joining a conflict.  With the rise in conflict, there was also a decline in affiliative behaviors: reconciliation, play, grooming, and physical proximity.  From these observations, \citet{Flack05study} concluded: 
\begin{quotation}
The extent to which policing is important to organizational robustness is surprising considering that actual policing behavior occurs relatively rarely.  This suggests that the simple presence of individuals responsible for conflict management can change the way group members are willing to interact with one another.
\end{quotation}

In my own work \citep{Frank96vary}, I developed the theoretical prediction with regard to policing when individuals vary in vigor or resources:
\begin{quotation}
Small variations in individual vigour or resources can lead to large variations in individual contributions to policing the group. Stronger individuals often invest all of their excess resources into policing, but weaker individuals do not contribute to group cohesion.
\end{quotation}
\citet{Flack05species} directly addressed this prediction:
\begin{quotation}
The primary finding of this study is that heterogeneities in power, by producing heterogeneities in the cost of conflict management for individuals, lead to heterogeneities in the tendency to police.
\end{quotation}
In pigtailed macaques, the well endowed make essentially all the investment in social control.

In a subsequent paper, \citet{Flack06primates} concluded that policing by dominant individuals plays a key role in group cohesion: 
\begin{quotation}
We observe that when policing is operational, group members build larger social networks characterized by greater partner diversity and increased potential for socially positive contagion and cooperation.  Without policing, high conflict frequency and severity leads to more conservative social interactions and a less integrated society.
\end{quotation}

\subsubsection{Domestication and Repression of Competition}

Humans have domesticated various animal and plant species for food production.  Ants began farming much earlier---about 50 million years ago \citep{Mueller02symbiosis}.  

Fungus growing ants collect plant material to feed their crops.  The ants weed their gardens to protect against fungal parasites that specialize in attacking fungus gardens.  The ants also grow specialized cultures of bacteria on their bodies in order to use the antibiotic secretions produced by their partner bacteria \citep{Zhang07ants}.  The antibiotics protect the fungal gardens from bacterial diseases.  

Domesticated fungal species were once wild, free living species.  New domesticates carry with them their own evolved tendencies for competition in local groups---their tragedy of the commons---in which such competition reduces the efficiency and productivity of the domesticates.  

I developed the general prediction that nonhuman masters gain by repressing competition among domesticates in order to elicit the most efficient domestic productivity \citep{Frank96lineages}.  In natural, unregulated situations, mixture between genetically unrelated strains often leads to greater competition between individuals and a decline in productivity.  Such competition develops between mixed lineages because mixture reduces the behavioral correlation between individuals.  Following this logic, the easiest way for ant farmers to reduce conflict between fungal domesticates would be to prevent mixing of fungal lineages in their gardens.  Homogeneous domesticates have high correlation in their cooperative tendencies, leading to an intrinsic tendency to reduce competition and enhance group productivity.  

Ants do in fact prevent mixing of domesticate fungal lineages \citep{Bot01ants,Mueller04Formicidae,Zhang07ants}.  When a newborn queen leaves her  birthplace to found a new colony, she brings with her the fungal lineage from her natal colony.  The fungi produce chemicals that inhibit growth by competing lineages; the ants spread those anti-competitor chemicals in their feces, which fertilize the growing fungal garden.

\subsection{Correlated Interests}

An individual may value payoff to another individual.  Such other regarding behavior arises commonly in biology from genetic relatedness.  For example, life depends on the regard a parent has for its offspring. 

In biology, we can tally the payoff to a parent for its various behaviors directed at offspring.  We first count the benefits of those behaviors for the offspring discounted by the genetic correlation between parent and offspring.  We then subtract off the direct cost of the behaviors for the parent.  This calculation yields Hamilton's rule, by which a behavior is favored when $rb-c>0$, where $r$ is the genetic correlation (or regression) between actor and recipient, $b$ is the benefit to the recipient, and $c$ is the cost to the actor---see Equation \eqref{hrOrig}.

The actor's valuation of the recipient in proportion to genetic correlation arises from the fact that transmission of strategies through time ultimately determines the evolutionary dynamics of behavior.  A recipient of a behavior carries the actor's deviation from the average strategy in proportion to the genetic correlation between actor and recipient.  

Other forces may sometimes affect the valuation of a recipient by the actor.  For example, the recipient may return a beneficial behavior at a later time \citep{Trivers71altruism}.  In this case, valuation of another arises as an investment in an expected future payback.  Most often, however, other regarding behavior in biology arises from genetic relatedness.

Parental altruism toward offspring occurs so widely in nature that I will not elaborate further on that case.  Instead, I focus on two interesting examples in which individuals weigh their comparative regard for different classes of genetic relatives.  The comparative aspect highlights the quantitative nature by which an actor regards the payoff to others.

\subsubsection{Competition for being the Queen}

In many social insect colonies, a newborn female may develop into a reproductive queen or a partially sterile worker \citep{Wilson71Societies}. A queen directly produces offspring; a worker helps to raise sisters and brothers.  A newborn female often would gain more by developing into a queen rather than a worker.  Here, I measure success by the biological standard of genetic contribution to future generations.  I tabulate total genetic success by the effect of a behavior on direct reproduction and on the reproduction of genetic relatives weighted by the genetic relatedness of actor to recipient.

The queen or the older workers usually control the fate of newborn females---development into either the queen or worker caste.  The elders control the caste of newborns by manipulating offspring size, by varying chemical stimulus, and by altering the provisioning of food.  The elders' control over caste represses competition between newborns over development into queens.

In a particular type of social bee, elders do not coerce the caste of newborns \citep{Wenseleers04bees}. Each newborn female may develop unconstrained into either a queen or a worker.  However, the number of available opportunities for newborn queens to lead a colony is limited.  Those who develop into queens compete with each other for those limited slots.  Only a small number of queens succeed in this competition; the losers die and do not contribute to the colony productivity.  

The competition between newborn queens causes a tragedy of the commons \citep{Wenseleers04bees}.  To increase group efficiency and productivity, the colony should avoid costly overproduction of queens, most of whom die in competition with their neighbors.  Those queens who die in competition could have developed into workers who contribute to common productivity.  However, individuals may gain by competing for the extra individual payoff of being a queen.  

The calculation of payoffs for being a worker or queen is more complex in this case than for the simple tragedy of the commons models I discussed earlier.  In this case, we must account for other regarding valuations ascribed to different kinds of genetic relatives.  In particular, we want to know how the behavioral choice of being a queen or a worker affects others in the colony, and how the actor, faced with the choice of alternative development, regards those who are affected by the choice.

A special aspect of bees, ants, and wasps arises from their asymmetric inheritance system.  A mother's unfertilized egg develops into a son; her fertilized egg develops into a daughter.  Queens mate with males and produce both unfertilized sons and fertilized daughters.  Workers do not mate, but can lay unfertilized sons.  

Workers will typically be offspring of the queen.  The workers help to rear the new eggs laid in the colony.  Most of those new eggs will be laid by the queen and will be the workers' sisters and brothers.  However, some of the workers will directly lay their own sons.  Thus, in rearing new eggs, a worker will also be helping to rear some of her sisters' sons---that is, her nephews.  

Here is a central point:  a female is more closely related to her nephew than to her brother because of the peculiar asymmetry in inheritance \citep{Hamilton72insects}.  So the ratio of nephews to brothers determines the valuation a worker gains from her effort to rear the eggs produced by the colony.  The more nephews produced directly by her sister-workers, the more a female values the eggs she helps to rear as a worker.

Now we return to the key behavioral choice.  A newborn female can become a queen and rear sons and daughters.  Or she can become a worker and rear sisters, brothers, and nephews.  The greater the ratio of nephews to brothers, the greater the valuation of being a worker via accounting for genetic regard for others.

\begin{figure}
\centerline{\includegraphics[width=3.5in]{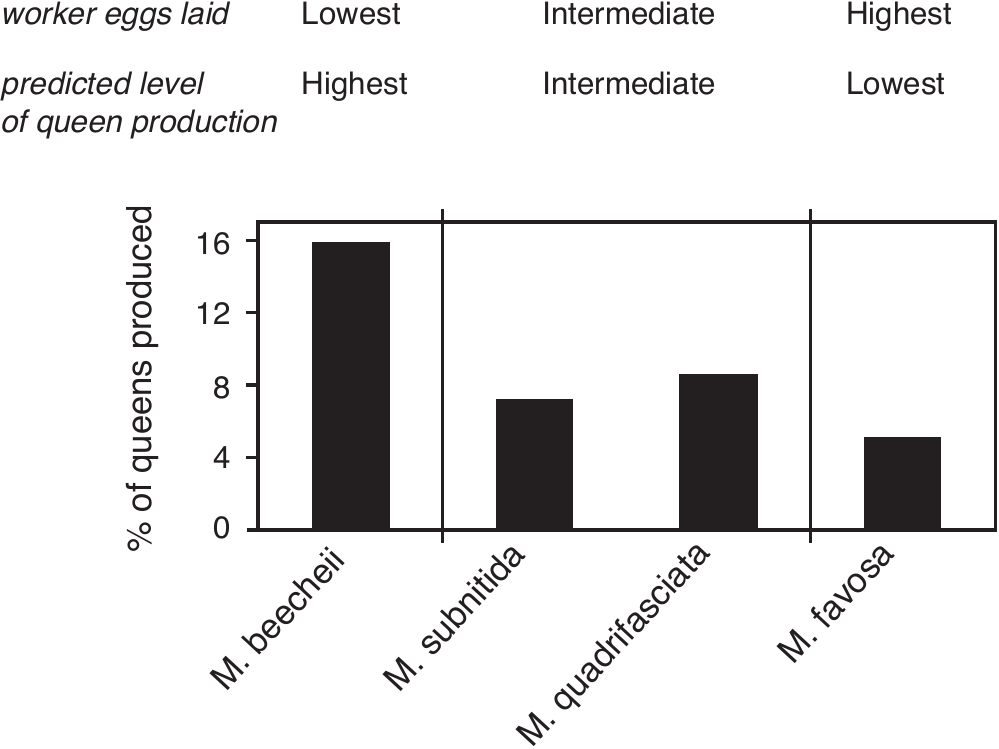}}
\caption{Fraction of newborn females who develop into queens.  More worker eggs laid means that a worker rears a higher fraction of more valuable nephews compared with brothers.  Theory predicts that as the fraction of more valuable nephews increases, more newborn females will develop into workers rather than queens, causing the percentage of queens produced to decline.  Data from four bee species of the genus \textit{Melipona\/} support the prediction. The percentage of queens produced in \textit{subnitida\/} and \textit{quadrifasciata\/} do not differ significantly, so those two species are lumped into a single intermediate category.  Redrawn from \citet{Wenseleers04bees}.}
\label{fig:beeWorkers} 
\end{figure}

Thus, we come to a simple prediction \citep{Wenseleers04bees}.  The fraction of females who develop into queens should decline as the amount of egg laying by workers rises.  To state the reasoning again: more egg laying by workers means that a worker will raise an increased ratio of nephews to brothers.  A worker values a nephew more highly than a brother.  As the fraction of nephews increases, the relative value of being a worker rises, and the relative gain of competing for a queenship drops compared to the expected gain of being a worker.  Figure \ref{fig:beeWorkers} shows data that support the prediction.  More egg laying by workers is associated with a lower fraction of newborn females developing into queens.

\subsubsection{Worker Valuation of Egg Production by Other Workers}I stated that, in the bees, ants, and wasps, a sister is more closely related to her nephew than to her brother.  This asymmetry occurs when a colony has a single queen who mates with only one male, as in the particular bees discussed in the previous example.  However, in some other species, a queen may mate several times, or there may be multiple queens.  The number of queens, the number of times a queen mates, and the level of inbreeding affect the asymmetry in relatedness of a worker to nephews and brothers \citep{Hamilton72insects}.

We can avoid complexity by considering a simple prediction.  When a worker is more related to the sons of other workers than to sons produced by the queen, she will allow other workers to produce sons without interference.  By contrast, when a worker is more related to the queen's sons than to sons produced by other workers, she will interfere with reproduction by other workers \citep{Wenseleers06theory}.

\begin{figure}
\centerline{\includegraphics[width=3.0in]{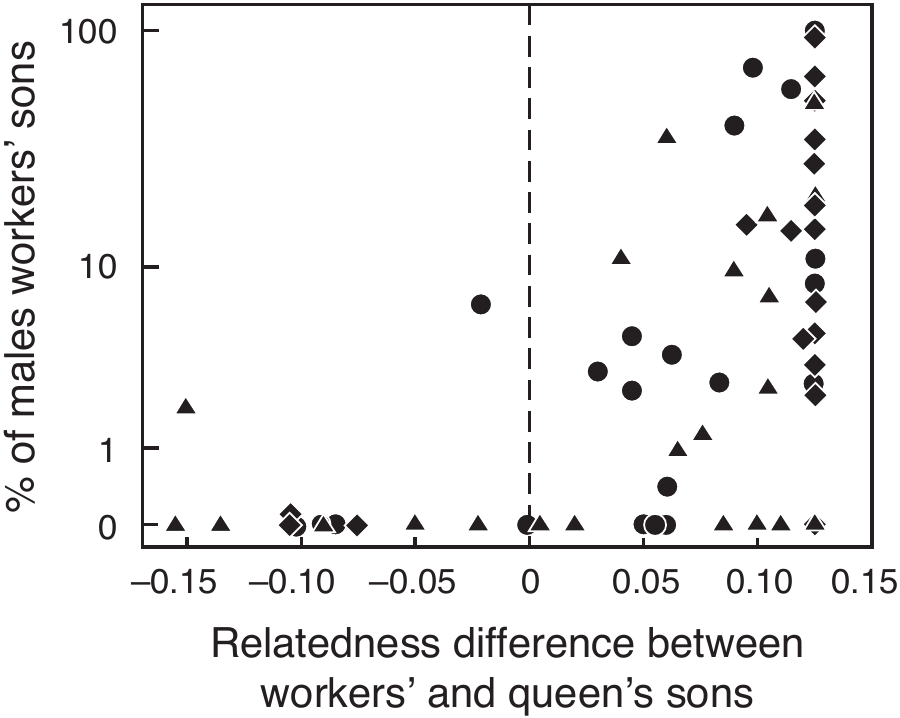}}
\caption{Relatedness asymmetry determines whether a worker allows other workers to reproduce. Negative values of the relatedness difference mean that a worker is more closely related to the queen's sons than to the workers' sons.  Positive values mean the opposite.  The height of each point shows the percentage of all males produced by a colony that derive from workers (scaled logarithmically).  The plot shows 90 different species of ants (circles), bees (squares), and wasps (triangles). Redrawn from \citet{Wenseleers06theory}.}
\label{fig:insectPolice} 
\end{figure}

Figure \ref{fig:insectPolice} supports the prediction that relatedness asymmetry determines worker behavior.  When workers are more related to other workers' sons than to the queen's sons, then the percentage of males produced by workers rises significantly above zero.  By contrast, when workers are more related to the queen's sons than to other workers' sons, production of sons by workers is almost always very close to zero.  In this case, the workers prevent other workers from producing sons by eating the eggs laid by workers.

These two examples from social insects show that the biological theory of other regarding valuation based on genetic kinship provides precise quantitative understanding of behavior.   

\subsection{Synergism and the Origin of Mutually Beneficial Behaviors}

The prior examples concerned behaviors that can be directly observed:  sex ratio, parasite virulence, or tolerance of egg laying by sister-workers.  By contrast, synergism concerns the origin of mutually beneficial behaviors at some time in the past.  

Mutually beneficial behaviors often require complementary specialization.  Such specialization frequently does not exist in advance of the cooperative venture \citep{Frank95symbiosis}.  So the first problem concerns how the mutually beneficial behavior got started and became sufficiently integrated to provide fully synergistic benefits.  

As parties come to depend on each other's complementary specializations, they may over time become mutually dependent.  If so, then a second problem concerns the irreversibility of synergistic behavior: neither party can succeed without the partnership.  Put another way, as each player becomes adjusted to the presence of the other, the other takes on the role of an essential part of the environment without which the individual cannot succeed.  Indeed, the integration can become so complete that it is hard to see the past history---in the present, the players have become so completely interdependent that we tend to view them as a unit.  

Much of the deep structure of life may have followed a path in which separate entities interacted synergistically and became mutually dependent. Synergism may be the most important of topics in the study of group integration, but it is also the most difficult of topics to analyze.   Past separation becomes hidden in present integration.

\subsubsection{The Origin of Integrated Individuals}

It is difficult to identify past synergism in current behaviors.  So I will start with a theoretical example.  The example concerns how the earliest kind of life may have become integrated into more complex cooperative groups.

Life depends on molecules that copy themselves.  Those molecules that replicate at the highest rate increase in abundance.  If the error rate in replication is sufficiently low, then a progeny molecule is mostly like its parent, and carries the same information that provided a replicative edge to its parent \citep{Eigen71macromolecules}.  

All of life became structured into cells early in history.  A cell contains the informational molecules that copy themselves.  Those informational molecules direct the biochemical physiology of the cell.  The physiology runs the program by which the cell acquires resources, protects itself against perturbation, and copies its informational molecules.

A cell is a complex cooperative consortium of multiple informational molecules, each informational component directing a part of the physiology needed to run the collaborative enterprise.  Because cells require complex integration of components, simple informational molecules that copy themselves must have preceded the earliest cells \citep{Eigen79SelfOrganization,Maynard79life}.   

How can different kinds of self interested molecules come to be associated in a mutually beneficial synergism?  

It is easy enough to imagine that if two different types express complementary information, then their interaction produces synergistic benefits.  But how did the two types come to express complementary information, if each initially evolved in isolation?  The problem turns on a threshold, as in the earlier theoretical section on synergism \citep{Frank95symbiosis}.  If both populations of alternative molecules express, on average, a level of complementary information above some threshold, synergism follows easily.  But the initial state is inevitably below the threshold.  

\begin{figure}
\centerline{\includegraphics[width=2.5in]{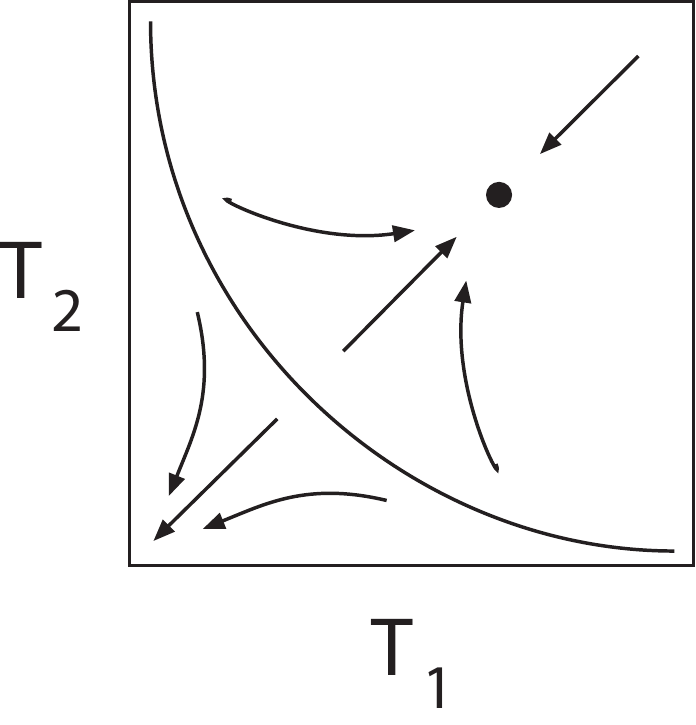}}
\caption{Thresold model for the origin and evolution of synergistic cooperation.  Individuals of population 1 have a trait, $T_1$, that enhances the reproductive rate of members of population 2, but the trait that benefits the partner also reduces the actor's own fitness.  Likewise, members of population 2 have a trait, $T_2$, that enhances the reproduction of individuals of population 1 at a cost to the actor.  Larger values of $T$ provide more benefit to the partner at a higher cost to the donor.  When both populations have low trait values, as would be expected when the partners first meet, natural selection continually pushes the traits to lower values.  If, however, the pair of traits is above a threshold upon first meeting, then cooperation can increase because of synergistic feedback.  Correlation in the traits between populations increases the probability that a particular group will have a pair of individuals above the threshold.  Such correlation may arise from spatial associations. From \citet{Frank95symbiosis}.}
\label{fig:threshold} 
\end{figure}

Figure \ref{fig:threshold} shows this threshold model for synergism.  Once both partners are above the threshold, mutually beneficial interactions strongly enhance the cooperative traits.  After enhancement by positive feedback, the partners may be investing heavily in traits that benefit each other in order to receive enhanced return benefits.  At that point, the partners may become fully dependent on each other for survival.  Once above the threshold, the likely path is: complementation and positive feedback $\rightarrow$ specialization $\rightarrow$ irreversible transition to a highly integrated and cooperative state \citep{Eigen79SelfOrganization}.

How can partners pass the threshold?  Suppose, initially, that the average traits of the two populations are below the threshold.  However, there is variability in each population.  So, by chance, some pairs of individuals will be above the threshold.  Those chance pairs will do better than average, because they gain the synergistic benefit from their positive feedback on each other.  Each will have more offspring than average, spreading the cooperative trait values in their populations.  However, if their progeny associate randomly with partners, then on average they will be investing highly in cooperation but will be matched with partners who do not reciprocate.  So those cooperative progeny will do less well than average, and no net progress in cooperative evolution ensues.  

Spatial associations may help \citep{Axelrod81cooperation,Nowak94cooperation}.  Suppose, by chance, that a pair of cooperators comes together.  They do well and leave more progeny than average.  If their progeny tend to associate rather than mix randomly, then the synergism continues, and the paired lineages of cooperators expand.  The spatial association, extended over time, allows the cooperative pairing to continue long enough to increase significantly, possibly pushing the average trait values over the threshold \citep{Frank94species,Frank95symbiosis}.  Once the average values pass the threshold, both populations rapidly enhance their synergistic traits, and the spatial associations are no longer required.  

Cells are membrane bound structures that naturally impose spatial associations \citep{Maynard95Evolution}.  Perhaps the spatial associations imposed by the early cells helped to push various biochemical synergisms over their cooperative thresholds.  As those thresholds were passed, the mutual dependence between molecules became fixed.  At that point, the biochemical integration became so deep that we would have a hard time recognizing the self interested histories behind the cohesive group.

\subsubsection{Irreversible Thresholds in the Social Evolution of Insect Colonies}

Positive feedback, historical change, and irreversibility may often play important roles in the evolution of complex and highly integrated groups.  In this section, I consider Bourke's \citeyearpar{Bourke99insects} interesting analysis of social insect colonies \citep{Frank03cooperation}.  Although Bourke's study transcends my simple models of synergism, his analysis does emphasize strongly the role of positive feedback and irreversible thresholds in cooperative evolution.

\citet{Bourke99insects} began by noting that, across different species of social insect, small colonies tend to have relatively little morphological differentiation between queens and workers. In addition, the workers have a relatively high degree of reproductive potential. By contrast, large colonies tend to have strong morphological differentiation between queens and workers and reduced reproductive potential of the workers.

\citet{Alexander91Rat} argued that in small colonies each worker has a significant probability of replacing the queen because there are relatively few competitors. By contrast, workers in large colonies have relatively little chance of succeeding to become queen. Thus, workers in large colonies are favored to reduce investment in reproductive potential and become more specialized for their worker roles. This reduction leads to strong morphological differentiation between workers and queens and low reproductive potential of workers. Absence of potential reproduction by workers reduces conflict between workers and other colony members, because the workers can enhance their fitness mostly by increasing the success of the colony.

\citet{Ratnieks92species} suggested that worker control of reproduction by other workers (policing) may be ineffective in small colonies. If there are few other workers, then a single worker may be able to dominate her neighbors and succeed in producing sons. As the number of workers rises, policing becomes more effective because a single worker cannot dominate the collective.

\citet{Bourke99insects} combined these ideas to argue that positive feedbacks occur between colony size, policing, reproductive potential of workers, and morphological differentiation between workers and queens. As colony size rises, policing becomes more effective, which favors reduced allocation to reproduction by workers. As workers concentrate more on their colony-productive roles, conflict subsides and the colony becomes more efficient. Greater efficiency may drive colonies to larger size, further specializing workers for nonreproductive tasks and aligning the interests of the workers with the interests of the colony.

\section{Historical Analogy}

The tension between conflict and cooperation of self interested actors runs deeply throughout the history of life.  The great puzzles turn on how cooperative and efficient groups arise solely through self interest.  The same basic forces seem to occur in both biology and in human affairs.  I first recap the biology, then comment on the analogy to humans.

In biology, the tragedy of the commons rules in the absence of any special force that promotes cooperation.  We see the tragedy of self interest everywhere:  in sex ratios, in parasite virulence, in bacterial secretion of resource-acquiring molecules.  In all cases, each member of a group would do better by promoting group cohesion and sharing in the benefits of greater group efficiency.  But, without some force that curbs the free expression of self interest, competition within the group ultimately plays against everyone's interests.

Several forces in biology have overcome the tragedy to promote group cohesion.  Correlated behaviors tie the success of actors together by matching behavior between partners and thus locking the success of the actor with its partner.  If an actor's success is tied with its neighbors, then the value of group efficiency can dominate the disruptive force of self interested competition.  In biology, correlated behavior most often arises by common genetics from shared ancestry---that is, by interactions between genetic kin.  But any force that induces correlation can have the same effect.

Repression of competition within groups ties the interests of each individual to the group.  With no opportunity to outcompete neighbors, each individual can gain only by promoting the efficiency and productivity of the group.  Any process that randomizes the share of group resources provided to each individual effectively represses any opportunity for competition.  The way that individuals divide their genetic material for transmission to offspring arose from a process of randomization, in which the probability that a particular gene passes to a child is determined randomly.  We call this process fair meiosis---the basis throughout much of life for sex, reproduction, inheritance, and individuality.  

Repression of competition may also be important at higher levels of social organization.  I discussed one study of a primate, in which dominant males police conflict in the group.  In the absence of those policing males, group cohesion deteriorated significantly.  Effective policing of competition may be difficult to achieve in many biological settings.  That difficulty explains why competition still rules much of life, and why in certain cases the ubiquitous force of competition may be overcome.

Self interested valuation of others' success arises naturally in biology through genetic kinship.  Observed patterns of behavior ultimately depend on the rate at which competing behaviors are transmitted into the future.  In this regard, an individual is shaped by natural selection to value the success of another in proportion to the correlation in their genetic tendency to pass the same behaviors on to future generations.

Regard for kin sets the foundation of social behavior.  It is the reason parents care for offspring, sterile honeybees raise their siblings, and nonreproductive skin cells die to promote the success of sperm or egg.  Kin correlations can shape behavior with great precision.  I illustrated that precision by the relative valuation a social insect worker places on the queen's sons versus the other workers' sons.  The class more highly valued switches depending on how many times the queen mates.  The workers switch their treatment accordingly.  When the workers are more closely correlated genetically with other workers' sons, they tolerate production of those sons without interference.  By contrast, when workers are more closely correlated genetically with the queen's sons, they destroy the sons produced by other workers.   

Finally, synergistic feedback between different aptitudes often provides benefit to both parties.  I discussed how such synergism likely played a key role in the earliest evolutionary history of life.  At some early stage, there must have been a consortium formed to produce the first cells.  That consortium arose between separate molecules, each originally designed to replicate itself but not to interact cooperatively with other molecules.  Some of those self interested replicators probably had synergistic biochemistry.  The positive feedbacks combined with spatial associations imposed by cellular boundaries set the first great cooperative transition of life.  I also discussed how such synergisms between complementary aptitudes shaped the historical trends to greater specialization and complexity in social insect colonies.

Correlated behaviors, repression of competition, other regarding valuation, and synergistic gains between different aptitudes rule conflict and cooperation throughout the history of life.  The potential analogies with human behavior are clear.  

But what is the value of such analogy between biology and human sociality?  I see two related benefits.  First, study of biology has greatly clarified the logic of self interest.  I have, in this paper, outlined a rich theory of conflict and cooperation that has succeeded well in explaining and in predicting diverse behaviors.  Many of those ideas from biology have arisen independently in the theory of games or in studies of human behavior.  But the biological theory has a firmer conceptual foundation and greater connection with observed phenomena.  

Second, analogies from nature suggest hypotheses about the forces that have shaped human societies.  For example, \citet{Alexander79Affairs,Alexander87Systems} has argued that many aspects of human morality turn on reducing competition between neighbors to promote group cohesion.  Alexander developed his hypothesis from close study of biology followed by analogy to human self interest.  

How useful are such analogies in forming hypotheses about human sociality?  Certainly, both large mistakes and great insights may follow from analogy.  Thus, one can reasonably defend both caution and boldness.  But caution cannot solve puzzles.  And many puzzles remain with regard to the forces that shape human cooperation and competition.

\section{Historical Consequence}

Historical analogy simply provides a source of ideas about how self interest plays out in human societies.  By contrast, historical consequence means that humans have been shaped by the same forces that have operated throughout biological history.  Such historical consequence still allows that humans are unique.  From a strictly biological perspective, humans have particular attributes that set us apart.  For example, other animals have culture and specialized cognitive abilities, but the great development of human culture and cognition define qualitative distinctions of human sociality.  

Thinking about historical consequence leads to obvious questions.  How strongly does genetic kinship shape behavior?  How much does learning and culture alter the dynamics of behavior?  How much bias has biological history built into the way we learn and transmit aspects of culture?  How much does a history of group against group competition align self interested tendencies with those of the group?

Biological history has had at least some consequential effects along these lines.  It simply does not make sense to suppose that history does not matter.  But we still do not know how to weigh various factors.  How should we proceed to learn more?

One approach is to consider the following question \citep{Alexander90evolve}:  What was the most important challenge to survival and reproduction that caused evolution to transform our ancestors from something like a chimpanzee into a modern human?  

To start, consider how biologists think about evolutionary transformation in response to challenges of survival and reproduction.  It is easier to see the structure of the argument if we begin with a nonhuman example.  I use kangaroos.  I describe the example in detail, because it is essential to understand the biological approach to analyzing evolutionary transformations.  After I finish with the kangaroo example, I apply the same logic to the evolutionary transformation of humans from their ape ancestors.

Most of the 63 species of kangaroo make their living on the ground.  Those ground dwelling species include the large well-known hoppers.  However, at least 10 species of kangaroo belong to a group that has become specialized for life in the trees \citep{Flannery96History}.   Adaptation for tree life led to several specialized characteristics.  

The key here is that a single broad challenge---moving from the ground into the trees---explains a wide array of evolutionary changes to deal with the challenge.  We will want to discuss what sort of challenge and what sort of changes characterize the evolutionary transformation of humans from ape ancestors.  But first, let use consider the kangaroos.  \citet{Diamond97kangaroos} has described the characteristic changes so well that I quote directly: 

\begin{quotation}
The lifestyle of arboreal tree-kangaroos required them to reverse millions of years of kangaroo evolution in many respects: saving weight by a 25 per cent reduction in muscle mass; developing long, strong, curved claws; big, powerful grasping forearms, and a rotator cuff in the shoulder (shared with humans but not with other kangaroos or most other mammals) to permit overhead use of the forearm; hind-feet that twist so that the soles can face each other to grasp a tree trunk; hair whorls to shed rain (also shared with humans); and a tail tufted with long fur in some species, used as a counterbalance in climbing and as a rudder in `flight'.

Those `flights' are actually jumps to the ground from a height of 20 metres or more in the canopy.  I know of no other big mammal that survives drops from such height $\ldots$ the animals' bones, muscles and ligaments must have become modified to withstand such shocks.

$\ldots$ Faced with a hard-to-digest, toxic, bulky leaf diet of low nutritional value, tree-kangaroos evolved a low metabolic rate.  They decrease rather than increase their metabolic rate at low ambient temperature; spend 90 per cent of their time `doing nothing' (that is, sitting and digesting); and have a complex stomach of several chambers, and regurgitate and rechew food, like cows chewing the cud.
\end{quotation}

The logic for tree kangaroos is simple.  They moved from the ground to the trees: that move defined the central evolutionary challenge.  Nearly all of the particular changes that separate tree kangaroos from those on the ground can be explained by evolutionary response to the key challenge.

We need to consider two steps in order to apply this same logic to humans.  First, what are the particular characteristics that separate humans from their ape ancestors?  This list of characteristics defines the changes that need to be explained.  People argue over the exact limits of the differences, for example, how much symbolic processing chimpanzees and gorillas can accomplish.  But in the end, these are more or less matters of fact that can be resolved by direct study.  

Nonetheless, any direct statement about humans is likely to be controversial.  Here is just one example of certain potentially unique traits of humans (Alexander 1990):

\begin{quotation}
Humans are the only mammal that lives in multi-male groups, in which confidence (likelihood) of paternity is high $\ldots$ and the males are both extensively and complexly parental and also extensively and complexly cooperative with one another (and in which, I speculate, the males with the highest confidence of paternity also tend to be the most cooperative.)
\end{quotation}

In the second step, we formulate a hypothesis: What is the central evolutionary challenge that allows us to make sense of the particular evolutionary transformations that separate humans from ape ancestors?  

Many theories have been proposed.  Several emphasize the social environment.  Again, I give just one brief account to indicate the way in which one may argue.  Alexander (1990) gives a full scholarly discussion of past work and presents his own views as follows:

\begin{quotation}
$\ldots$ At some point in their evolution humans obviously began to cooperate to compete $\ldots$ this intergroup competition becoming increasingly elaborate, direct, and continuous until it achieved the ubiquity with which it has been exhibited in modern humans throughout recorded history across the entire face of the earth $\ldots$ This unique kind of within-species balance-of-power race---involving, eventually, virtually all levels, or group sizes, within societies---would be a perpetual or unending one $\ldots$ in which rapidly appearing differences in culture and technology could become significant unbalancers that could accelerate the process even further.  It could be termed a case of ``runaway social selection''$\ldots$ [calling] for adversarial and competing groups of humans to be central in creating the environment of brain and psyche selection.  Unprecedented levels of cooperation within groups could thereby be generated, as well as unprecedented kinds of between-group adversarial relationships.
\end{quotation}

These points establish the problem of historical consequence.  One may follow this problem in many directions.  But, whatever direction, we can be sure that the concepts of self interested cooperation will play an important role in framing the key evolutionary challenge and the particular historical consequences for distinctly human characteristics.

It is, of course, possible that multiple factors explain the human transformation---different human characteristics may have different, unrelated explanations.  But I follow Michael Ghiselin \citeyearpar{Ghiselin69Method} in his analysis of Darwin's greatness:  above all, Darwin succeeded by his stubborn belief that a few simple processes explain much of the great complexity and variety of life.  To discover a simple explanation, one must assume that a simple explanation is possible.

\begin{acknowledgement}
National Institute of General Medical Sciences MIDAS Program grant U01-GM-76499 supports my research.  
\end{acknowledgement}

\bibliography{Refs/Biblio/coop1}
\bibliographystyle{spbasic}
\end{document}